\documentclass[prl,showpacs,twocolumn, superscriptaddress]{revtex4}%
\usepackage{amsfonts}
\usepackage{amsmath}
\usepackage{amssymb}
\usepackage{graphicx}

\begin{document}

\title{Measuring many-body effects in carbon nanotubes \\ with a scanning tunneling microscope}

\author{H. Lin}
\affiliation{Laboratoire Mat\'{e}riaux et Ph\'{e}nom\`{e}nes Quantiques, Universit\'{e} Paris Diderot, CNRS, UMR7162, 10 rue Alice Domon et L\'{e}onie Duquet, 75205 Paris Cedex 13, France}
\affiliation{Laboratoire d'Etude des Microsctructures, ONERA-CNRS,
BP72, 92322 Ch\^{a}tillon, France}
\author{J. Lagoute}
\author{V. Repain}
\author{C. Chacon}
\author{Y.~Girard}
\affiliation{Laboratoire Mat\'{e}riaux et Ph\'{e}nom\`{e}nes Quantiques, Universit\'{e} Paris Diderot, CNRS, UMR7162, 10 rue Alice Domon et L\'{e}onie Duquet, 75205 Paris Cedex 13, France}
\author{J.-S. Lauret}
\affiliation{Laboratoire de Photonique Quantique et Mol\'{e}culaire, Institut d'Alembert, Ecole Normale Sup\'{e}rieure de Cachan, 94235 Cachan Cedex, France}
\author{F. Ducastelle}
\author{A. Loiseau}
\affiliation{Laboratoire d'Etude des Microsctructures, ONERA-CNRS,
BP72, 92322 Ch\^{a}tillon, France}
\author{S. Rousset}
\affiliation{Laboratoire Mat\'{e}riaux et Ph\'{e}nom\`{e}nes Quantiques, Universit\'{e} Paris Diderot, CNRS, UMR7162, 10 rue Alice Domon et L\'{e}onie Duquet, 75205 Paris Cedex 13, France}
\date{\today}

\begin{abstract}

Electron-electron interactions and excitons in carbon nanotubes are locally measured by combining Scanning tunneling spectroscopy and optical absorption in bundles of nanotubes. The largest gap deduced from measurements at the top of the bundle is found to be related to the intrinsic quasi-particle gap. From the difference with optical transitions, we deduced exciton binding energies of 0.4 eV for the gap and 0.7 eV for the second Van Hove singularity. This provides the first experimental evidence of substrate-induced gap renormalization on SWNTs.

\end{abstract}


\maketitle


The electronic properties of single-walled carbon nanotubes (SWNT)
have attracted wide interest since their discovery. 
The tunability from metallic to semiconducting character through a structural
control makes them promising candidates for the development
of molecular electronics. 
It is therefore crucial to understand and
control their electronic properties down to the atomic scale.
Scanning tunneling microscopy (STM) and spectroscopy (STS) 
are powerful tools for the local investigation of 1D band structure of
SWNTs and of its link to the atomic structure
\cite{Wilder1998,Odom1998,Venema2000}. STS experiments have
been successfully interpreted within a simple tight-binding
zone-folding scheme which shows a density of states dominated by a
series of Van-Hove singularities \cite{Hamada1992,Saito1992}.
The energy separation between the first two singularities
on each side of the Fermi level is denoted $E_{11}^{S}$ for
semiconducting tubes and $E_{11}^{M}$ for metallic tubes.
The simplest tight-binding model  gives a relation between
the electronic gap
$E_{11}^{S}$ and the tube diameter $d,
E_{11}^{S}=2\gamma_{0}a_{c-c}/d$ where $a_{c-c}$ is the
carbon-carbon distance (1.42 \AA) and $\gamma_{0}$ is the hopping
integral between first neighbours, ranging from 2.5 \cite{Kim1999}
to 2.9 eV \cite{Venema2000a}.
Although the tight-binding model captures the main features of the
electronic structure of nanotubes  \cite{Loiseau2006}, many-body
effects play a role in 1D systems and significantly
modify the energy of the electronic transitions. Two main
contributions should be taken into account: self-energy effects which
modify the single-particle excitation spectrum and, in the case of
optical transitions, excitonic electron-hole interactions
\cite{Ando1997,Kane2003,Dresselhaus2007,Ando2009}.

The major role of excitons in the optical transition of nanotubes has been
demonstrated by using two-photon absorption experiments
\cite{Wang2005,Dukovic2005,Maultzsch2005}. After this
finding, the interpretation of subband transitions measured in
optical experiments has been revisited. Although a parameterized
tight-binding model could explain the so-called Kataura plot, it turned out
that a complete description should include many-body effects. Theoretical
discussions show that self-energy and excitonic effects tend to
cancel each other, both effects being of the same order of
magnitude, in the range of 0.5 to 1 eV. The overall result is a weak blue
shift effect \cite{Kane2003,Kane2004,Jiang2007}.

In the same way, scanning tunneling microscopy results on SWNTs must
be revisited to some extend. STM experiments are
not expected to be sensitive to the electron-hole interaction and
one expects the gap measured by STM to be larger than the optical
one, the difference corresponding to the exciton binding energy.
However, to perform STS measurements, the tubes are usually placed on a metallic substrate.
Charge transfer between the metallic substrate and the tube shifts
the Fermi level of the tube \cite{Janssen2002,Vitali2006}.
In the case of molecules, the renormalization of the electronic gap
by substrate-induced charging effects in STS measurements has been
studied both theoretically and experimentally
\cite{Hesper1997,Lu2004,Sau2008} and a similar effect is expected
for SWNTs \cite{Kane2004} although detailed experimental data are still lacking.

In this Letter, we provide experimental evidence
of this substrate-induced gap renormalization on SWNTs.
STS measurements show a systematic variation of the band gap measured on
semi-conducting tubes, depending on whether the tube is on the top of
a bundle or in direct contact with the metallic substrate. A
single-particle model for $E_{11}^{S}$ is not sufficient to account
for the experimental data. On the other hand a quasi-particle model
combined with an analysis of substrate-induced screening effects
allows us to propose a consistent interpretation of our
experiments. The unscreened gap deduced from
STS measurements is later compared with the optical absorption
spectrum providing us with a unique tool for
determining the exciton binding energies.


STM/STS measurements were performed using a low temperature (5K) STM operating under UHV conditions
($10^{-10}$ mbar). The nanotubes were
synthesized by arc discharge technique and deposited from a
dispersion in alcohol after sonication onto a single-crystalline Au
(111) surface previously cleaned in UHV by repeated
cycles of ion argon sputtering and annealing at 800 K. The samples
were dried in air and introduced into the UHV system. Although the
preparation was aimed to avoid bundling of nanotubes, bundles are
still observed after the deposition. All measurements were performed
with tungsten tips. Optical absorption spectra were recorded with a spectrophotometer (lambda 900 Perkin-Elmer) on the same solution
deposited onto a glass substrate.


The electronic gap of a nanotube is principally  determined by its
diameter. It is therefore crucial to have a precise knowledge of the
diameter distribution of the nanotubes. Although STM allows us to
attain atomic resolution, tip profile effects lead to a poor
estimation of the tube diameters. To overcome that problem, we
combined the STM investigation with transmission electron microscopy (TEM) analysis which, in contrast,
can provide more accurate values of these diameters. Using a gaussian
fit we obtain a statistical distribution
centered at 1.4 nm with a FWHM of 0.2 nm. This
distribution has been projected onto the Kataura plots (see Fig. 1) to deduce the
induced gap distributions.

\begin{figure}[!ht]
    \centering
    \includegraphics[width=85mm]{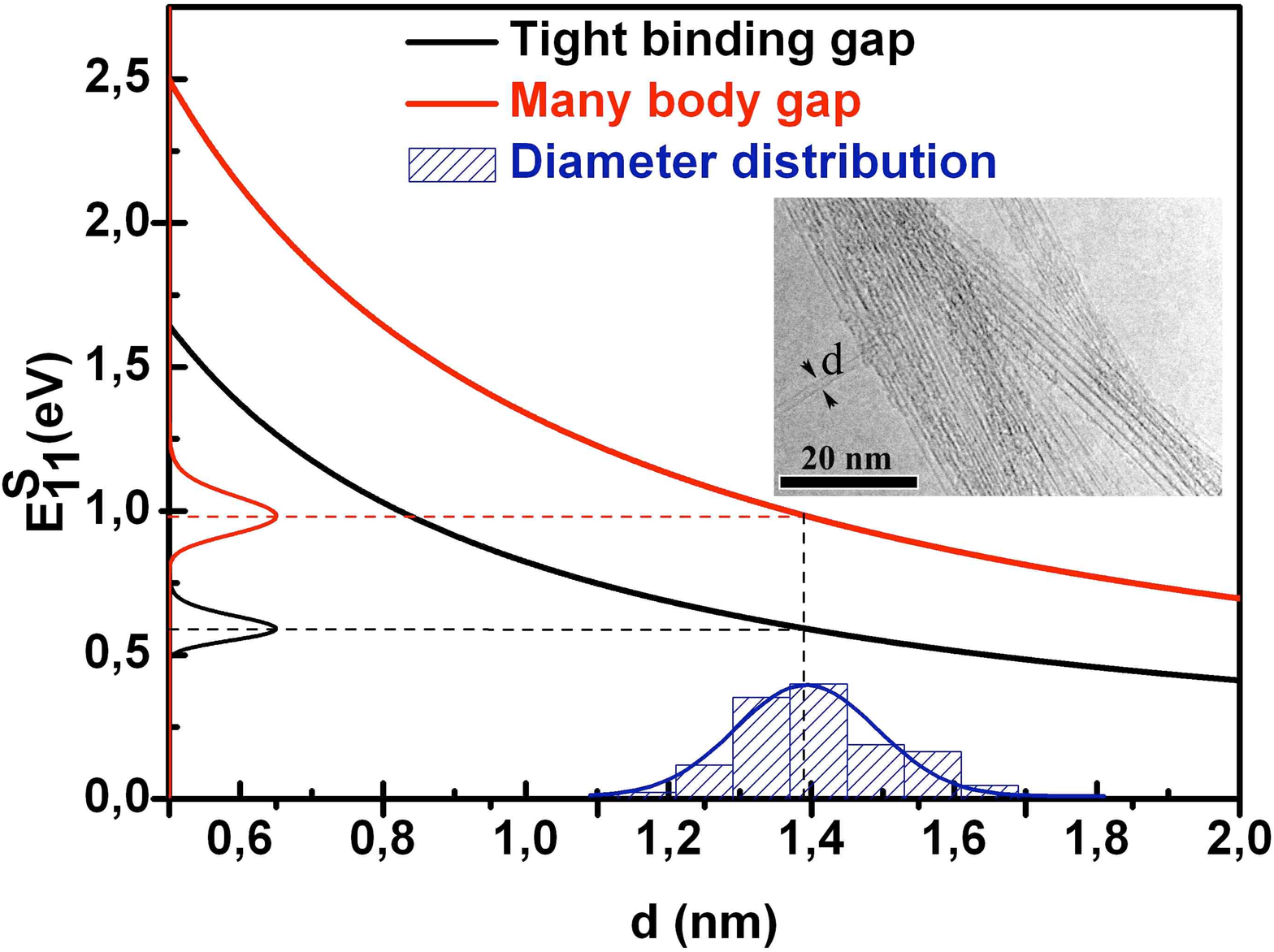}
    \caption{(Color online) Statistic distribution of nanotubes diameter measured
    on TEM images of 50 SWNTs (Philips CM20, 200 kV). The inset presents an example of a
    TEM image. The lower and upper curves
    are the $E_{11}^{s}$ gap using the single-particle tight-binding model or the many body values
    \cite{Dukovic2005}, respectively. The corresponding gap
     distributions are shown on the vertical axis.}
    \label{STM}
\end{figure}

Fig.\ref{STM} displays first the
plot deduced from the simplest linearized tight-binding model (lower curve),
$E^{s}_{11}=2\gamma_{0}a_{c-c}/d$. Using $\gamma_{0}=2.9$ eV, this
plot predicts a gap distribution centred at 0.59 eV (see Fig.\
\ref{STM}). The second plot is deduced from two-photon
experiments \cite{Dukovic2005} where the single-quasiparticle gap is
expressed by $E^{s}_{11}=0.34/d+1.11/(d+0.11)$ (upper curve in Fig.\ \ref{STM}).
We now obtain a gap distribution centered at 0.98 eV. The difference
between the two curves corresponds to the many-body
self-energy contributions. The upper curve will be referred as ``many-body'' gap, even if it should be kept in
mind that excitonic effects are not included \cite{Dukovic2005}.

\begin{figure}[!ht]
    \centering
     \includegraphics[width=85mm]{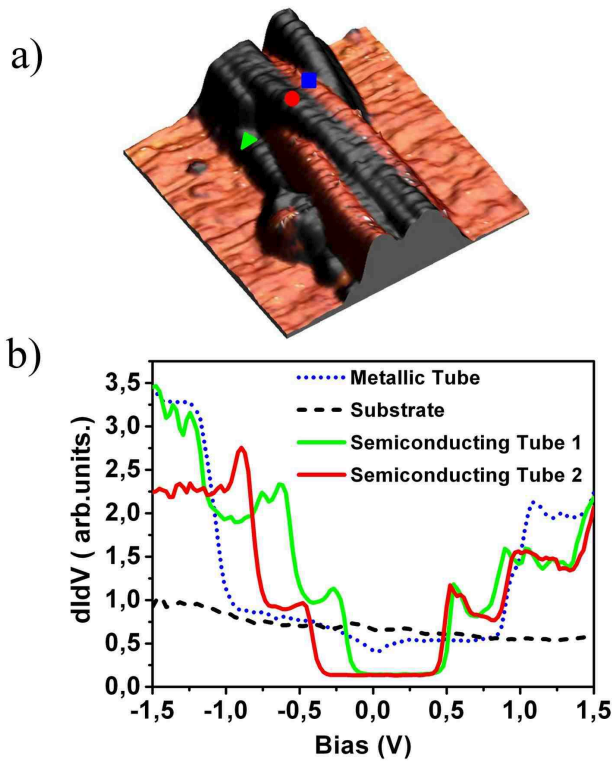}
    \caption{(Color online) STM/STS of a bundle of tubes on the metallic surface.
a) 3D view of the topographic image (40x40 nm$^2$) with a color
scale corresponding to the differential conductance image at 12 mV
(black for low value, light gray (yellow online) for high value).
b) dI/dV spectra taken at positions indicated in a).}
    \label{gap}
\end{figure}

Because of the narrowness of the diameter distribution, the simple
tight-binding and the many-body gap distributions do no overlap. It
is then crucial to understand which one corresponds to the value
measured in a STM experiment. We argue hereafter that, to some extend,
and depending on the tube-substrate separation, both are measured.
After deposition on the Au(111) substrate, most nanotubes arrange
into bundles. The differential conductance image at 12 mV (see Fig.\
\ref{gap}(a)) shows the distribution of the local density of state
around the Fermi level and allows us to separate clearly the
metallic (bright) and semi-conducting tubes (black) in the image.
Fig.\ \ref{gap}(b) shows typical $dI/dV$ spectra measured on the
substrate and on the tubes. On the gold substrate, the differential
conductance has a constant non-zero value as expected (black dashed
curve). On the nanotubes, spectra are dominated by Van-Hove singularities.
On the metallic tube, the conductance between the two first
singularities is close to that of the substrate, except for a dip
around the Fermi level (dotted curve). This pseudo-gap was observed
for all the metallic tubes investigated. It is known to be due to
curvature effects or intertube interactions and has already been
observed \cite{Ouyang2001}.

\begin{figure}
    \centering
    \includegraphics[width=85mm]{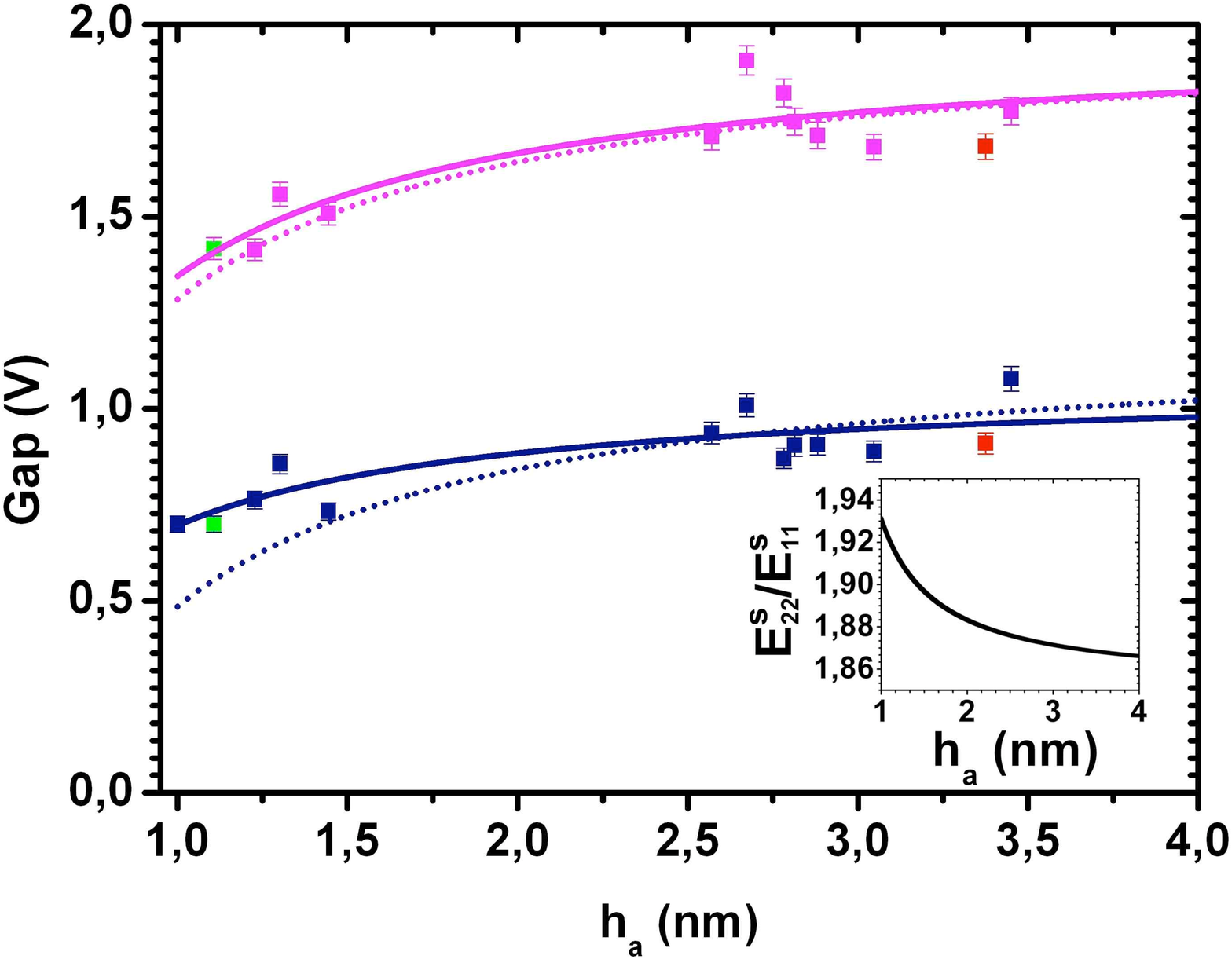}
     \caption{(Color online) Gap as a function of the apparent height (h$_{a}$) of the tubes.
The squares are the experimental data. The dotted lines are fits
using image charge model. The solid lines are fits using effective parameters $C_1$ and $C_2$. The inset shows the
$E_{22}^{S}$/$E_{11}^{S}$ ratio. The green and red squares
correspond to the tubes 1 and 2 of Fig.\ \ref{gap}, respectively.
The error bars are estimated from the diameter dispersion
projected onto the tight-binding gap curve.}
     \label{ratio}
\end{figure}

Two spectra of semi-conducting tubes showing the first two Van-Hove
singularities are presented in Fig.\ \ref{gap}(b). The peak
positions are not symmetric with respect to the Fermi level, which
is due to tube-substrate charge transfer
\cite{Wilder1998,Odom1998,Venema2000}. The energy subband
separations $E^{S}_{11}$ are measured at the point of maximum slope
in the peak.  The values when the tube directly contacts the
substrate (tube 1) or is at the top of the bundle (tube 2) are equal
to 0.70 eV (green spectrum) and 0.91 eV (red spectrum) respectively
(Fig. 2(b)). It is worth noticing that we did not measure
similar variations of the $E^{M}_{11}$ transition for metallic
tubes. As shown in Fig.\ref{STM},  the dispersion in the gap values
of semiconductor tubes cannot be accounted for by the width of the
diameter distribution within one or the other model.
On the other hand the gap of the tube
contacting the metal compares well with the gap given by the
single-particle model, while the gap of the tube separated from the
substrate by the bundle lies within the distribution estimated
by the ``many-body'' Kataura plot. The transition between these two
situations can be understood considering the potential image effect
which decreases when the tube-sample distance increases.
Within this model the $E_{11}^{S}$ transition of a nanotube adsorbed
on a metal corresponds to the gap of the free nanotube (
including electron-electron interactions) reduced by the screening
energy due to the image charge in the metal $C_0e^{2}/2D$, where D
is the tube-substrate distance and
$C_0$=$1/4\pi\varepsilon_{0}$ \cite{Hesper1997,Sau2008}. It is
therefore expected that a nanotube separated from the metallic
substrate has a larger gap than a nanotube in contact. In
our experiment the bundle plays the role of a spacer which offers
the possibility to measure the gap of semiconducting nanotubes at
various tube-substrate separations. The largest gap measured at high
height should then correspond to the genuine intrinsic
quasi-particle gap of the nanotubes. To verify this hypothesis, we
present in Fig.\ \ref{ratio} the first and second subband energy
separation as a function of the apparent height ($h_{a}$) with respect to the gold substrate 
of several nanotubes as determined from STM images at 1 V. We see that for both the first
and second singularities, the energy separation ($E_{11}^{S}$ and
$E_{22}^{S}$, respectively) tends to increase with the apparent
height. The image charge model ($E=E_{0}-C_{0}e^{2}/(2h_{a})$) fits
reasonably well the experimental data (Fig. 3, dotted curves).

This first approach however does not consider the effective
dielectric constant of the tube environment, neither the fact that
the apparent height is not exactly equal to the tube-substrate
distance. For these reasons we provide empirical fits using
$E=E_{0}-C_{1}e^{2}/(2h_{a})$, where $C_{1}=0.52C_0$ for the
$E_{11}$ gap and $E=E_{0}-C_{2}e^{2}/(2h_{a})$ with $C_{2}=0.89C_0$
for the $E_{22}$ separation (solid lines in Fig.\
\ref{ratio}). Note that $C_{1} \ne C_{2}$ what is ascribed to many-body effects, as discussed later. When $h_{a}$$\rightarrow\infty$,
$E_{0}$ for the first and the second singularities tends to 1.1 eV
and 2.0 eV, respectively. From our analysis, these values should therefore be equal to the
quasi-particle gaps. $E_{11}^{s}=1.1$ eV is close to the gap
deduced by the many-body relation with d=1.4 nm (0.98 eV), and
definitely  larger than the tight-binding calculation (0.59 eV).

\begin{figure}
    \centering
    \includegraphics[width=85mm]{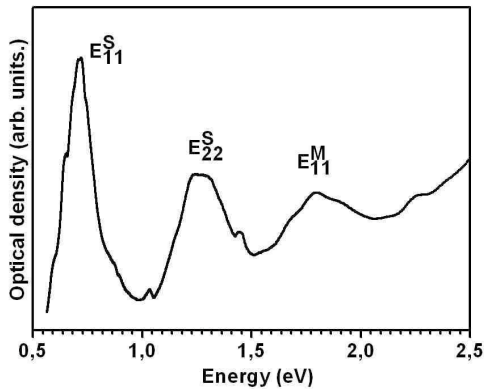}
     \caption{(Color online) Optical absorption of SWNTs. $E_{11}^{S}$, $E_{22}^{S}$
and $E_{11}^{M}$ are equal to 0.72,1.26, and 1.79 eV respectively.
The $E_{22}^{S}$/$E_{11}^{S}$ ratio is equal to 1.75.}
    \label{absorption}
\end{figure}

We now compare STS data with optical absorption spectrum
of the same nanotubes sample displayed in Fig.\ \ref{absorption}.
It is clear that the optical transitions occur at lower
energies than those obtained in STS on the top of
bundles (Fig.\ \ref{ratio}). The energy difference $\Delta E$
between the optical transitions and the tunneling gap in the limit of infinite $h_{a}$ is then a
measure of the exciton binding energies. We obtain $\Delta
E_{11}^{s}$=0.4 eV and $\Delta E_{22}^{s}$=0.7 eV. The
$E_{22}^{s}$ states have larger excitonic binding energy than
$E_{11}^{s}$ states, which is consistent with theoretical
calculations \cite{Ando2004,Jiang2007}. The $\Delta E_{11}^{s}$
value obtained here is close to the binding energy determined in
previous optical measurements where a law $\Delta E_{11}^{s} \simeq
0.34/d $ has been derived \cite{Dukovic2005}. Therefore 
combining STS with optical spectroscopy we can estimate the exciton binding energy of carbon
nanotubes.

Finally, we discuss the $E_{22}^{S}$/$E_{11}^{S}$ ratio measurement.
This ratio is generally found experimentally smaller than 2 (about
1.8), the value provided by the simple linearized tight-binding
model: this is the ``ratio problem'' \cite{Kane2003}. It is now clear that the
main explanation for this deviation is the effect of
electron-electron interactions  \cite{Kane2003,Dresselhaus2007},
even if the relative contributions of self-energy and excitonic
effects are not precisely known. Calculations show also fairly large
fluctuations of this ratio as a function of the chirality.
In our STS experiments, sensitive to self-energy effects only, a
$E_{22}^{S}$/$E_{11}^{S}$ ratio close to 2 corresponds well to the
left hand part of the curve in the inset of Fig.\ref{ratio}, where
the self-energy is largely compensated by the screening effect due
to the image potential. As the tube-substrate distance increases,
self-energy effects increase and the $E_{22}^{S}$/$E_{11}^{S}$ ratio
decreases gradually down to 1.85. Previous experimental data
reported by Venema \emph{et al.} \cite{Venema2000a} give a ratio close to 2
, but their tubes lie on the metal, and according
to our analysis, they correspond to a situation where self-energy
effects are screened out.  Considering now our optical absorption
measurements, a lower $E_{22}^{S}$/$E_{11}^{S}$ ratio about 1.75 is
found. Our data indicate then that the gap ratio decreases when
the screening of electron-electron interactions decreases.


To summarize, we used STS measurements to compare local spectroscopy
of carbon nanotubes lying on a metallic substrate with the
spectroscopy of nanotubes on top of bundles. We showed that the gap
of a semiconducting tube increases when it is separated from the
metal by a bundle, the gap variation being of the order of what is
expected within an image charge model where self-energy effects are
screened. The experimental conditions  allowed us to observe a continuous
transition from the almost totally screened case, to an almost
intrinsic gap. By combining these
data with optical absorption measurement, the exciton binding energies could then be estimated.
In the case of 1.4 nm diameter nanotubes a mean value of 0.4 eV
and 0.7 eV was found, for excitons corresponding to the
$E_{11}^{S}$ and $E_{22}^{S}$ gaps. We believe that these
measurements will open promising ways of determining
locally the exciton binding energy of carbon nanotubes.

\begin{acknowledgments}
This study has been supported by the European Contract STREP ``BCN''
Nanotubes 30007654-OTP25763, by a grant of CNano IdF ``SAMBA'' and
by the ANR project ``CEDONA'' of the PNANO programme (ANR-07-NANO-007\_02). We gratefully acknowledge L. Henrard and P. Hermet
for fruitful discussions.
\end{acknowledgments}


\begin{thebibliography}{25}
\expandafter\ifx\csname natexlab\endcsname\relax\def\natexlab#1{#1}\fi
\expandafter\ifx\csname bibnamefont\endcsname\relax
  \def\bibnamefont#1{#1}\fi
\expandafter\ifx\csname bibfnamefont\endcsname\relax
  \def\bibfnamefont#1{#1}\fi
\expandafter\ifx\csname citenamefont\endcsname\relax
  \def\citenamefont#1{#1}\fi
\expandafter\ifx\csname url\endcsname\relax
  \def\url#1{\texttt{#1}}\fi
\expandafter\ifx\csname urlprefix\endcsname\relax\def\urlprefix{URL }\fi
\providecommand{\bibinfo}[2]{#2}
\providecommand{\eprint}[2][]{\url{#2}}

\bibitem[{\citenamefont{Wilder et~al.}(1998)\citenamefont{Wilder, Venema,
  Rinzler, Smalley, and Dekker}}]{Wilder1998}
\bibinfo{author}{\bibfnamefont{J.~W.~G.} \bibnamefont{Wilder}},
  \bibinfo{author}{\bibfnamefont{L.~C.} \bibnamefont{Venema}},
  \bibinfo{author}{\bibfnamefont{A.~G.} \bibnamefont{Rinzler}},
  \bibinfo{author}{\bibfnamefont{R.~E.} \bibnamefont{Smalley}},
  \bibnamefont{and} \bibinfo{author}{\bibfnamefont{C.}~\bibnamefont{Dekker}},
  \bibinfo{journal}{Nature} \textbf{\bibinfo{volume}{391}}, \bibinfo{pages}{59}
  (\bibinfo{year}{1998}).

\bibitem[{\citenamefont{Odom et~al.}(1998)\citenamefont{Odom, Huang, Kim, and
  Lieber}}]{Odom1998}
\bibinfo{author}{\bibfnamefont{T.~W.} \bibnamefont{Odom}},
  \bibinfo{author}{\bibfnamefont{J.-L.} \bibnamefont{Huang}},
  \bibinfo{author}{\bibfnamefont{P.}~\bibnamefont{Kim}}, \bibnamefont{and}
  \bibinfo{author}{\bibfnamefont{C.~M.} \bibnamefont{Lieber}},
  \bibinfo{journal}{Nature} \textbf{\bibinfo{volume}{391}}, \bibinfo{pages}{62}
  (\bibinfo{year}{1998}).

\bibitem[{\citenamefont{Venema et~al.}(2000{\natexlab{a}})\citenamefont{Venema,
  Meunier, Lambin, and Dekker}}]{Venema2000}
\bibinfo{author}{\bibfnamefont{L.~C.} \bibnamefont{Venema}},
  \bibinfo{author}{\bibfnamefont{V.}~\bibnamefont{Meunier}},
  \bibinfo{author}{\bibfnamefont{P.}~\bibnamefont{Lambin}}, \bibnamefont{and}
  \bibinfo{author}{\bibfnamefont{C.}~\bibnamefont{Dekker}},
  \bibinfo{journal}{Phys. Rev. B} \textbf{\bibinfo{volume}{61}},
  \bibinfo{pages}{2991} (\bibinfo{year}{2000}{\natexlab{a}}).

\bibitem[{\citenamefont{Hamada et~al.}(1992)\citenamefont{Hamada, Sawada, and
  Oshiyama}}]{Hamada1992}
\bibinfo{author}{\bibfnamefont{N.}~\bibnamefont{Hamada}},
  \bibinfo{author}{\bibfnamefont{S.-i.} \bibnamefont{Sawada}},
  \bibnamefont{and} \bibinfo{author}{\bibfnamefont{A.}~\bibnamefont{Oshiyama}},
  \bibinfo{journal}{Phys. Rev. Lett.} \textbf{\bibinfo{volume}{68}},
  \bibinfo{pages}{1579} (\bibinfo{year}{1992}).

\bibitem[{\citenamefont{Saito et~al.}(1992)\citenamefont{Saito, Fujita,
  Dresselhaus, and Dresselhaus}}]{Saito1992}
\bibinfo{author}{\bibfnamefont{R.}~\bibnamefont{Saito}},
  \bibinfo{author}{\bibfnamefont{M.}~\bibnamefont{Fujita}},
  \bibinfo{author}{\bibfnamefont{G.}~\bibnamefont{Dresselhaus}},
  \bibnamefont{and} \bibinfo{author}{\bibfnamefont{M.~S.}
  \bibnamefont{Dresselhaus}}, \bibinfo{journal}{Phys. Rev. B}
  \textbf{\bibinfo{volume}{46}}, \bibinfo{pages}{1804} (\bibinfo{year}{1992}).

\bibitem[{\citenamefont{Kim et~al.}(1999)\citenamefont{Kim, Odom, Huang, and
  Lieber}}]{Kim1999}
\bibinfo{author}{\bibfnamefont{P.}~\bibnamefont{Kim}},
  \bibinfo{author}{\bibfnamefont{T.~W.} \bibnamefont{Odom}},
  \bibinfo{author}{\bibfnamefont{J.-L.} \bibnamefont{Huang}}, \bibnamefont{and}
  \bibinfo{author}{\bibfnamefont{C.~M.} \bibnamefont{Lieber}},
  \bibinfo{journal}{Phys. Rev. Lett.} \textbf{\bibinfo{volume}{82}},
  \bibinfo{pages}{1225} (\bibinfo{year}{1999}).

\bibitem[{\citenamefont{Venema et~al.}(2000{\natexlab{b}})\citenamefont{Venema,
  Janssen, Buitelaar, Wild\"'oer, Lemay, Kouwenhoven, and Dekker}}]{Venema2000a}
\bibinfo{author}{\bibfnamefont{L.~C.} \bibnamefont{Venema}},
  \bibinfo{author}{\bibfnamefont{J.~W.} \bibnamefont{Janssen}},
  \bibinfo{author}{\bibfnamefont{M.~R.} \bibnamefont{Buitelaar}},
  \bibinfo{author}{\bibfnamefont{J.~W.~G.} \bibnamefont{Wild\"oer}},
  \bibinfo{author}{\bibfnamefont{S.~G.} \bibnamefont{Lemay}},
  \bibinfo{author}{\bibfnamefont{L.~P.} \bibnamefont{Kouwenhoven}},
  \bibnamefont{and} \bibinfo{author}{\bibfnamefont{C.}~\bibnamefont{Dekker}},
  \bibinfo{journal}{Phys. Rev. B} \textbf{\bibinfo{volume}{62}},
  \bibinfo{pages}{5238} (\bibinfo{year}{2000}{\natexlab{b}}).

\bibitem[{\citenamefont{Loiseau et~al.}(2006)\citenamefont{Loiseau, Launois,
  Petit, Roche, and Salvetat}}]{Loiseau2006}
\bibinfo{editor}{\bibfnamefont{A.}~\bibnamefont{Loiseau}},
  \bibinfo{editor}{\bibfnamefont{P.}~\bibnamefont{Launois}},
  \bibinfo{editor}{\bibfnamefont{P.}~\bibnamefont{Petit}},
  \bibinfo{editor}{\bibfnamefont{S.}~\bibnamefont{Roche}}, \bibnamefont{and}
  \bibinfo{editor}{\bibfnamefont{J.-P.} \bibnamefont{Salvetat}}, eds.,
  \emph{\bibinfo{title}{Understanding carbon nanotubes}}
  (\bibinfo{publisher}{Springer}, \bibinfo{year}{2006}).

\bibitem[{\citenamefont{Ando}(1997)}]{Ando1997}
\bibinfo{author}{\bibfnamefont{T.}~\bibnamefont{Ando}}, \bibinfo{journal}{J.
  Phys. Soc. Jpn} \textbf{\bibinfo{volume}{66}}, \bibinfo{pages}{1066}
  (\bibinfo{year}{1997}).

\bibitem[{\citenamefont{Kane and Mele}(2003)}]{Kane2003}
\bibinfo{author}{\bibfnamefont{C.~L.} \bibnamefont{Kane}} \bibnamefont{and}
  \bibinfo{author}{\bibfnamefont{E.~J.} \bibnamefont{Mele}},
  \bibinfo{journal}{Phys. Rev. Lett.} \textbf{\bibinfo{volume}{90}},
  \bibinfo{pages}{207401} (\bibinfo{year}{2003}).

\bibitem[{\citenamefont{Dresselhaus et~al.}(2007)\citenamefont{Dresselhaus,
  Dresselhaus, Saito, and Jorio}}]{Dresselhaus2007}
\bibinfo{author}{\bibfnamefont{M.~S.} \bibnamefont{Dresselhaus}},
  \bibinfo{author}{\bibfnamefont{G.}~\bibnamefont{Dresselhaus}},
  \bibinfo{author}{\bibfnamefont{R.}~\bibnamefont{Saito}}, \bibnamefont{and}
  \bibinfo{author}{\bibfnamefont{A.}~\bibnamefont{Jorio}},
  \bibinfo{journal}{Annu. Rev. Phys. Chem.} \textbf{\bibinfo{volume}{58}},
  \bibinfo{pages}{719} (\bibinfo{year}{2007}).

\bibitem[{\citenamefont{Ando and Seiji}(2009)}]{Ando2009}
\bibinfo{author}{\bibfnamefont{T.}~\bibnamefont{Ando}} \bibnamefont{and}
  \bibinfo{author}{\bibfnamefont{U.}~\bibnamefont{Seiji}},
  \bibinfo{journal}{Phys. Stat. Sol. (c)} \textbf{\bibinfo{volume}{6}},
  \bibinfo{pages}{173} (\bibinfo{year}{2009}).

\bibitem[{\citenamefont{Wang et~al.}(2005)\citenamefont{Wang, Dukovic, Brus,
  and Heinz}}]{Wang2005}
\bibinfo{author}{\bibfnamefont{F.}~\bibnamefont{Wang}},
  \bibinfo{author}{\bibfnamefont{G.}~\bibnamefont{Dukovic}},
  \bibinfo{author}{\bibfnamefont{L.~E.} \bibnamefont{Brus}}, \bibnamefont{and}
  \bibinfo{author}{\bibfnamefont{T.~F.} \bibnamefont{Heinz}},
  \bibinfo{journal}{Science} \textbf{\bibinfo{volume}{308}},
  \bibinfo{pages}{838} (\bibinfo{year}{2005}).

\bibitem[{\citenamefont{Dukovic et~al.}(2005)\citenamefont{Dukovic, Wang, Song,
  Sfeir, Heinz, and Brus}}]{Dukovic2005}
\bibinfo{author}{\bibfnamefont{G.}~\bibnamefont{Dukovic}},
  \bibinfo{author}{\bibfnamefont{F.}~\bibnamefont{Wang}},
  \bibinfo{author}{\bibfnamefont{D.}~\bibnamefont{Song}},
  \bibinfo{author}{\bibfnamefont{M.~Y.} \bibnamefont{Sfeir}},
  \bibinfo{author}{\bibfnamefont{T.~F.} \bibnamefont{Heinz}}, \bibnamefont{and}
  \bibinfo{author}{\bibfnamefont{L.~E.} \bibnamefont{Brus}},
  \bibinfo{journal}{Nano Lett.} \textbf{\bibinfo{volume}{5}},
  \bibinfo{pages}{2314} (\bibinfo{year}{2005}).

\bibitem[{\citenamefont{Maultzsch et~al.}(2005)\citenamefont{Maultzsch,
  Pomraenke, Reich, Chang, Prezzi, Ruini, Molinari, Strano, Thomsen, and
  Lienau}}]{Maultzsch2005}
\bibinfo{author}{\bibfnamefont{J.}~\bibnamefont{Maultzsch}},
  \bibinfo{author}{\bibfnamefont{R.}~\bibnamefont{Pomraenke}},
  \bibinfo{author}{\bibfnamefont{S.}~\bibnamefont{Reich}},
  \bibinfo{author}{\bibfnamefont{E.}~\bibnamefont{Chang}},
  \bibinfo{author}{\bibfnamefont{D.}~\bibnamefont{Prezzi}},
  \bibinfo{author}{\bibfnamefont{A.}~\bibnamefont{Ruini}},
  \bibinfo{author}{\bibfnamefont{E.}~\bibnamefont{Molinari}},
  \bibinfo{author}{\bibfnamefont{M.~S.} \bibnamefont{Strano}},
  \bibinfo{author}{\bibfnamefont{C.}~\bibnamefont{Thomsen}}, \bibnamefont{and}
  \bibinfo{author}{\bibfnamefont{C.}~\bibnamefont{Lienau}},
  \bibinfo{journal}{Phys. Rev. B} \textbf{\bibinfo{volume}{72}},
  \bibinfo{pages}{241402(R)} (\bibinfo{year}{2005}).


\bibitem[{\citenamefont{Kane and Mele}(2004)}]{Kane2004}
\bibinfo{author}{\bibfnamefont{C.~L.} \bibnamefont{Kane}} \bibnamefont{and}
  \bibinfo{author}{\bibfnamefont{E.~J.} \bibnamefont{Mele}},
  \bibinfo{journal}{Phys. Rev. Lett.} \textbf{\bibinfo{volume}{93}},
  \bibinfo{pages}{197402} (\bibinfo{year}{2004}).

\bibitem[{\citenamefont{Jiang et~al.}(2007)\citenamefont{Jiang, Saito,
  Samsonidze, Jorio, Chou, Dresselhaus, and Dresselhaus}}]{Jiang2007}
\bibinfo{author}{\bibfnamefont{J.}~\bibnamefont{Jiang}},
  \bibinfo{author}{\bibfnamefont{R.}~\bibnamefont{Saito}},
  \bibinfo{author}{\bibfnamefont{G.~G.} \bibnamefont{Samsonidze}},
  \bibinfo{author}{\bibfnamefont{A.}~\bibnamefont{Jorio}},
  \bibinfo{author}{\bibfnamefont{S.~G.} \bibnamefont{Chou}},
  \bibinfo{author}{\bibfnamefont{G.}~\bibnamefont{Dresselhaus}},
  \bibnamefont{and} \bibinfo{author}{\bibfnamefont{M.~S.}
  \bibnamefont{Dresselhaus}}, \bibinfo{journal}{Phys. Rev. B}
  \textbf{\bibinfo{volume}{75}}, \bibinfo{pages}{035407}
  (\bibinfo{year}{2007}).

\bibitem[{\citenamefont{Janssen et~al.}(2002)\citenamefont{Janssen, Lemay,
  Kouwenhoven, and Dekker}}]{Janssen2002}
\bibinfo{author}{\bibfnamefont{J.~W.} \bibnamefont{Janssen}},
  \bibinfo{author}{\bibfnamefont{S.~G.} \bibnamefont{Lemay}},
  \bibinfo{author}{\bibfnamefont{L.~P.} \bibnamefont{Kouwenhoven}},
  \bibnamefont{and} \bibinfo{author}{\bibfnamefont{C.}~\bibnamefont{Dekker}},
  \bibinfo{journal}{Phys. Rev. B} \textbf{\bibinfo{volume}{65}},
  \bibinfo{pages}{115423} (\bibinfo{year}{2002}).

\bibitem[{\citenamefont{Vitali et~al.}(2006)\citenamefont{Vitali, Burghard,
  Wahl, Schneider, and Kern}}]{Vitali2006}
\bibinfo{author}{\bibfnamefont{L.}~\bibnamefont{Vitali}},
  \bibinfo{author}{\bibfnamefont{M.}~\bibnamefont{Burghard}},
  \bibinfo{author}{\bibfnamefont{P.}~\bibnamefont{Wahl}},
  \bibinfo{author}{\bibfnamefont{M.~A.} \bibnamefont{Schneider}},
  \bibnamefont{and} \bibinfo{author}{\bibfnamefont{K.}~\bibnamefont{Kern}},
  \bibinfo{journal}{Phys. Rev. Lett.} \textbf{\bibinfo{volume}{96}},
  \bibinfo{pages}{086804} (\bibinfo{year}{2006}).

\bibitem[{\citenamefont{Hesper et~al.}(1997)\citenamefont{Hesper, Tjeng, and
  Sawatzky}}]{Hesper1997}
\bibinfo{author}{\bibfnamefont{R.}~\bibnamefont{Hesper}},
  \bibinfo{author}{\bibfnamefont{L.~H.} \bibnamefont{Tjeng}}, \bibnamefont{and}
  \bibinfo{author}{\bibfnamefont{G.~A.} \bibnamefont{Sawatzky}},
  \bibinfo{journal}{Europhys. Lett.} \textbf{\bibinfo{volume}{40}},
  \bibinfo{pages}{177} (\bibinfo{year}{1997}).

\bibitem[{\citenamefont{Lu et~al.}(2004)\citenamefont{Lu, Grobis, Khoo, Louie,
  and Crommie}}]{Lu2004}
\bibinfo{author}{\bibfnamefont{X.}~\bibnamefont{Lu}},
  \bibinfo{author}{\bibfnamefont{M.}~\bibnamefont{Grobis}},
  \bibinfo{author}{\bibfnamefont{K.~H.} \bibnamefont{Khoo}},
  \bibinfo{author}{\bibfnamefont{S.~G.} \bibnamefont{Louie}}, \bibnamefont{and}
  \bibinfo{author}{\bibfnamefont{M.~F.} \bibnamefont{Crommie}},
  \bibinfo{journal}{Phys. Rev. B} \textbf{\bibinfo{volume}{70}},
  \bibinfo{pages}{115418} (\bibinfo{year}{2004}).

\bibitem[{\citenamefont{Sau et~al.}(2008)\citenamefont{Sau, Neaton, Choi,
  Louie, and Cohen}}]{Sau2008}
\bibinfo{author}{\bibfnamefont{J.~D.} \bibnamefont{Sau}},
  \bibinfo{author}{\bibfnamefont{J.~B.} \bibnamefont{Neaton}},
  \bibinfo{author}{\bibfnamefont{H.~J.} \bibnamefont{Choi}},
  \bibinfo{author}{\bibfnamefont{S.~G.} \bibnamefont{Louie}}, \bibnamefont{and}
  \bibinfo{author}{\bibfnamefont{M.~L.} \bibnamefont{Cohen}},
  \bibinfo{journal}{Phys. Rev. Lett.} \textbf{\bibinfo{volume}{101}},
  \bibinfo{pages}{026804} (\bibinfo{year}{2008}).

\bibitem[{\citenamefont{Ouyang et~al.}(2001)\citenamefont{Ouyang, Huang,
  Cheung, and Lieber}}]{Ouyang2001}
\bibinfo{author}{\bibfnamefont{M.}~\bibnamefont{Ouyang}},
  \bibinfo{author}{\bibfnamefont{J.-L.} \bibnamefont{Huang}},
  \bibinfo{author}{\bibfnamefont{C.~L.} \bibnamefont{Cheung}},
  \bibnamefont{and} \bibinfo{author}{\bibfnamefont{C.~M.}
  \bibnamefont{Lieber}}, \bibinfo{journal}{Science}
  \textbf{\bibinfo{volume}{292}}, \bibinfo{pages}{702} (\bibinfo{year}{2001}).

\bibitem[{\citenamefont{Ando}(2004)}]{Ando2004}
\bibinfo{author}{\bibfnamefont{T.}~\bibnamefont{Ando}}, \bibinfo{journal}{J.
  Phys. Soc. Jpn} \textbf{\bibinfo{volume}{73}}, \bibinfo{pages}{3351}
  (\bibinfo{year}{2004}).

\end{thebibliography}

\end{document}